\documentclass[fleqn, useAMS, usenatbib]{mn2e}

\usepackage{times}
\usepackage{epsfig, amssymb}
\usepackage{graphics}
\usepackage{graphicx}
\usepackage{eufrak}
\usepackage{mathrsfs}
\usepackage{eucal}
\usepackage{amsmath}
\usepackage{fontenc}
\usepackage{aas_macros} 

\def\etal{et al.~}  
\def\ie{{i.e.,\ }}
\def\galform{{\sc galform }}

\newcommand{\ergsec}	{\ifmmode \mathrm{\,erg~s}^{-1} \else \,erg~s$^{-1}$\fi}
\newcommand{\yr}	{\ifmmode \mathrm{yr} \else yr\fi}
\newcommand{\mpc}	{\ifmmode \,\mathrm{Mpc}^{-3} \else \,Mpc$^{-3}$\fi}
\newcommand{\Msun}	{\ifmmode \,\mathrm M_{\odot} \else $\,\mathrm M_{\odot}$\fi}
\newcommand{\Mbh}	{\ifmmode M_{\mathrm{BH}} \else $M_{\mathrm{BH}}$\fi}
\newcommand{\Mseed}	{\ifmmode M_{\mathrm{BH,seed}} \else $M_{\mathrm{BH,seed}}$\fi}
\newcommand{\Mbulge}	{\ifmmode M_{\mathrm{Bulge}} \else $M_{\mathrm{Bulge}}$\fi}
\newcommand{\Mhalo}	{\ifmmode M_{\mathrm{Halo}} \else $M_{\mathrm{Halo}}$\fi}
\newcommand{\Mhaloeff}	{\ifmmode M_{\mathrm{Halo,eff}} \else $M_{\mathrm{Halo,eff}}$\fi}
\newcommand{\Medd}	{\ifmmode \dot{M}_{\mathrm{Edd}} \else $\dot{M}_{\mathrm{Edd}}$\fi}
\newcommand{\Lbol}	{\ifmmode L_{\mathrm{bol}} \else $L_{\mathrm{bol}}$\fi}
\newcommand{\Ledd}	{\ifmmode L_{\mathrm{Edd}} \else $L_{\mathrm{Edd}}$\fi}
\newcommand{\Lcool}	{\ifmmode L_{\mathrm{cool}} \else $L_{\mathrm{cool}}$\fi}
\newcommand{\Lsun}	{\ifmmode L_{\odot} \else $L_{\odot}$\fi}
\newcommand{\Lsx}	{\ifmmode L_{\mathrm{SX}} \else $L_{\mathrm{SX}}$\fi}
\newcommand{\Mbj}	{\ifmmode M_{b_{\rm J}} \else $M_{b_{\rm J}}$\fi}
\newcommand{\bj}		{\ifmmode b_{\rm J} \else $b_{\rm J}$\fi}
\newcommand{\Lhx}	{\ifmmode L_{\mathrm{HX}} \else $L_{\mathrm{HX}}$\fi}
\newcommand{\ledd}	{\ifmmode \lambda_{\mathrm{Edd}} \else $\lambda_{\mathrm{Edd}}~$\fi}
\newcommand{\lgledd}	{\ifmmode \lambda_{\mathrm{Edd}} \else $\lambda_{\mathrm{Edd}}$\fi}
\newcommand{\rhobh}	{\ifmmode \rho_{\mathrm{BH}} \else $\rho_{\mathrm{BH}}$\fi}
\newcommand{\mdot}	{\ifmmode \dot{m} \else $\dot{m}$\fi}
\newcommand{\fhalo}	{\ifmmode f_{\mathrm{Halo}}^{\mathrm{act}} \else $f_{\mathrm{Halo}}^{\mathrm{act}}$\fi}
\newcommand{\fvis}	{\ifmmode f_{\mathrm{vis}} \else $f_{\mathrm{vis}}$\fi}
\newcommand{\fobsc}	{\ifmmode f_{\mathrm{obsc}} \else $f_{\mathrm{obsc}}$\fi}
\newcommand{\fq}		{\ifmmode f_{\mathrm{q}} \else $f_{\mathrm{q}}$\fi}
\newcommand{\fbh}	{\ifmmode F_{\mathrm{BH}} \else $F_{\mathrm{BH}}$\fi}

\title[The halo environment of luminous quasars]{The most luminous quasars do not live in the most massive dark matter haloes at any redshift}
\author[Fanidakis \etal]{N. Fanidakis$^{1}$\thanks{E-mail: fanidakis@mpia.de}, A.~V.~Macci\`o$^{1}$, 
C.~M.~Baugh$^{2}$, 
C.~G.~Lacey$^{2}$, 
C.~S.~Frenk$^{2}$\\
$^1$ Max-Planck-Institute for Astronomy, K\"onigstuhl 17, D-69117 Heidelberg, Germany.\\
$^2$ Institute for Computational Cosmology, Department of Physics, University of Durham, Science Laboratories, South Road, Durham, DH1 3LE, UK.\\}

\begin{document}

\maketitle
\label{firstpage}
\begin{abstract}
Quasars represent the brightest Active Galactic Nuclei (AGN) in the Universe and are thought to indicate the location of prodigiously growing Black Holes (BHs), with luminosities as high as $10^{48}\ergsec$. It is often expected though that such an extremely energetic process will take place in the most massive bound structures in the dark matter (DM) distribution. We show that in contrast to this expectation, in a galaxy formation model which includes AGN feedback, quasars are predicted to live in average DM halo environments with typical masses of a few times $10^{12}\Msun$. This fundamental prediction arises from the fact that quasar activity (\ie BH accretion with luminosity greater than $10^{46}\ergsec$) is inhibited in DM haloes where AGN feedback operates. The galaxy hosts of quasars in our simulations are identified with over massive (in gas and stars) spheroidal galaxies, in which BH accretion is triggered via a galaxy merger or secular processes. We further show that the $z=0$ descendants of high redshift ($z\gtrsim6$) QSOs span a wide range of morphologies, galaxy and halo masses. The $z\sim6$ BHs typically grow only by a modest factor by the present day. Remarkably, high redshift QSOs never inhabit the largest DM haloes at that time and their descendants are very seldom found in the most massive haloes at $z=0$. We also show that  observationally it is very likely to find an enhancement in the abundance of galaxies around quasars at $z\sim5$. However, these enhancements are considerably weaker compared to the overdensities expected at the extreme peaks of the DM distribution. Thus, it is very unlikely that a quasar detected in the $z\gtrsim5$ Universe pinpoints the location of the progenitors of superclusters in the local Universe. 

\end{abstract}
\begin{keywords}
 cosmology:dark matter -- cosmology:large-scale structure of Universe -- cosmology:theory -- galaxies:haloes -- galaxies:quasars
\end{keywords}

\section{Introduction}
The clustering of dark matter (DM) haloes is a well studied problem, since it is exclusively determined by the nature of the DM particle, gravity and the expansion history of the Universe. Numerical simulations and analytical models show that there is a strong dependence of clustering on mass, with higher mass haloes being more clustered than lower mass haloes \citep{Sheth1999}. On the other hand, the clustering of galaxies depends strongly on the physics of galaxy formation (gas cooling, star formation, feedback) and does not exactly map the distribution of DM. As a consequence, galaxies are biased tracers of the DM distribution (\citealt{Kaiser1984}; \citealt{Bardeen1986}; \citealt{Cole1989}; \citealt{MoWhite1996}). However, despite its complexity, there is strong dependence of galaxy clustering on luminosity, which implies that more luminous galaxies live in more massive haloes than less luminous galaxies (\citealt{Norberg2001}; \citealt{Zehavi2005}). In this context, quasars have gained immense popularity as galaxy tracers of the most massive haloes at high redshifts, because they are extremely bright and are observed at great distances. Consequently, if quasars live in the most massive haloes it is expected that those detected at high redshifts should directly probe the early growth of these structures. Yet, it remains unclear whether or not bright quasars do indeed reside in the extremes of the dark matter distribution.

In the low-$z$ Universe ($z\lesssim2$), clustering studies of quasars in large surveys such as the Sloan Digital Sky Survey (SDSS) and the 2dF, suggest that the typical DM halo mass of luminous quasars is $(2-3)\times10^{12}h^{-1}\Msun$ (\citealt{Wake2004}; \citealt{Porciani2004}; \citealt{Croom2005}; \citealt{Ross2009}, see also \citealt{Shanks2011}). This is considerably lower than the mass of the largest haloes in place at that redshift (which typically is of the order of $10^{14}-10^{15}\Msun$) and independent of quasar luminosity \citep[see however][]{Shen2012}. Therefore, quasars in the low-$z$ Universe reside in average regions of the DM distribution. This picture is supported by semi-analytical studies, in which quasar activity is usually driven by galaxy-galaxy mergers (\citealt{Bonoli2009}; \citealt{Bonoli2010}, see also \citealt{Marulli2008}). The model by \citet{Bonoli2009}, which is built upon the galaxy formation model of \citet{DeLucia2007}, predicts an average halo mass of $10^{12}-10^{13}\Msun$ for quasars. Quasar activity in more massive haloes is typically inhibited due to the suppression of gas cooling by AGN feedback.

In the high-$z$ Universe ($z\gtrsim2$) the picture is not clear. The clustering of $z\geq2.9$ quasars in the SDSS indicates a minimum halo mass of $(3-6)\times10^{12}h^{-1}\Msun$, slightly more massive than that of their lower-$z$ counterparts \citep{Shen2007}. The strong clustering of quasars in their sample implies that quasars in the high-$z$ Universe are tracers of highly biased massive haloes. These predictions sparked numerous studies of the environment of high-$z$ quasars. In these studies, the environment of quasars is usually probed by estimating the abundance of emission line galaxies such as H$\alpha$ galaxies, Ly-$\alpha$ emitters  (LAEs) or Lyman Break Galaxies (LBGs) around the quasar. A higher abundance of galaxies compared to the field typically indicates an overdensity, which then implies that the host halo is relatively overmassive. At such high redshifts, these structures could be collapsing in today's clusters and therefore, it is likely that the quasar under investigation pinpoints the direct location of a protocluster. Interestingly, there is an ambiguity regarding the conclusions of the different studies in the literature. Even though a certain number of studies suggest that quasars indeed trace massive structures (\citealt{Cantalupo2012}; \citealt{Swinbank2012}) an equal number claim that the environment of quasars is average (\citealt{Francis2004}; \citealt{Kashikawa2007}, see also \citealt{Swinbank2012}). 

The quest to find overdensities around quasars becomes of particular interest when the most distant ($z\geq5$) quasars are considered. The relatively high luminosities of these objects indicate that the mass of the BHs powering accretion is close to $10^{9}\Msun$, already at $z=7$ (Mortlock \etal 2011). This implies that the BHs in the $z\sim6$ quasars have grown in environments where the existence of abundant cold gas is favoured. It is appealing then to associate these environments with the most massive structures of DM, where gas cooling is expected to be prodigious. For this reason, it has been assumed that the most distant quasars reside at the peaks of the DM distribution \citep{Fan2003} and therefore, trace the location of the progenitors of today's superclusters. This assumption has become the norm in theoretical studies of the evolution of the DM distribution and galaxies, where quasars are typically associated with the most massive DM haloes in the early universe (\citealt{Springel2005}; \citealt{Overzier2009}; \citealt{Capak2011}; \citealt{Angulo2012}). 

Observationally, there have been several studies that support this scenario, usually by probing overdensities of faint $i_{775}$-dropout (\citealt{Stiavelli2005}; \citealt{Zheng2006}) or sub-mm galaxies \citep{Priddey2008} in the fields of $z\sim6$ quasars. However, \citet{Willott2005} performed deep optical imaging of three $6.2<z<6.5$ quasar fields and found no evidence of $i'$-band dropout overdensities \citep[see also][]{Carilli2004}. Similarly, \citet{Kim2009} observed $i_{775}$-dropouts in five $z\sim6$ SDSS quasars fields and found that only 2 show any evidence of an overdensity. Finally, in a more recent study, Banados \etal (in prep.) searched for LBGs and LAEs in the field of a $z=5.7$ quasar, probing LAEs in a narrow redshift range of $\Delta z\simeq 0.1$. The authors show that the LBG and LAE abundances are consistent with those found in random fields and therefore the quasar does not reside in an overdensity. Thus, the picture emerging for the $z\geq5$ quasars is also not clear.

Here we present a study of the DM environment of quasars by employing the semi-analytic model \texttt{GALFORM}. In this model, the formation and evolution of galaxies and BHs is fully coupled, and modelled consistently within the hierarchical clustering of the DM distribution (\citealt{Fanidakis2011}; \citealt{Fanidakis2012}). The aim of the study is to shed light on the typical halo mass of quasars in the low and high-$z$ Universe and to provide a physical framework within which the aforementioned observations can be explained. The paper is organised as follows. In Section~2 we briefly describe the main points of the model used in this analysis. In Section~3 we present the predictions of the model for the correlation between AGN luminosity and host halo mass, and demonstrate how quasars inhabit average DM environments. In Section~4 we explore the environmental dependence of the brightest quasars in the early universe and trace their descendants to $z=0$. In the same section we also make predictions for the expected number of galaxies around quasars in order to reconcile the observations. Finally, we complete our analysis by summarising our findings in Section~5. The cosmology adopted in our simulations is similar to the best constrains on the cosmological parameters from the analysis of the seven-year data release from WMAP \citep[WMAP7,][]{Komatsu2011}. Throughout this paper we choose: $\Omega_{\mathrm{m}}=0.227$, $\Omega_{\mathrm{b}}=0.045$, $\Omega_{\mathrm{\Lambda}}=0.728$ and $\sigma_8=0.81$\footnote{$\Omega_{\mathrm{m}}$, $\Omega_{\mathrm{b}}$ and $\Omega_{\mathrm{\Lambda}}$ express the present density of the baryonic, total matter and dark energy components of the Universe relative to the critical density ($\rho_{\mathrm{crit}}=3H^2/8\pi G$). $\sigma_8$ measures the rms mass fluctuations in spheres of radius $8~h^{-1}\mathrm{Mpc}$ linearly extrapolated to the present day.}. We set $h=0.7$ for all galaxy properties that we calculate.

\section{The model}

To tackle the key questions of galaxy formation several techniques have been devised over the past two decades. Among the most prominent is semi-analytical modelling \citep[see][for a review]{Baugh2006, Benson2010}. Semi-analytical models combine the strength of direct N-body simulations of the DM density field with the flexibility of a set of coupled differential equations that describe the physical processes that govern galaxy formation and evolution. The former approach has the advantage of being computationally inexpensive and therefore, ideal for exploring the BH parameter and model (adding new physics) space. Among the most prominent semi-analytical models is \galform \citep{Cole2000}.

\galform takes into account in a self consistent way all the main processes involved in galaxy formation: i) formation and evolution of DM haloes in the $\Lambda$CDM cosmology, ii) gas cooling and disc formation in DM haloes, iii) star formation, supernova feedback and chemical enrichment, iv) BH growth and AGN feedback, v) and formation of bulges during galactic disc instabilities and galaxy mergers. The model has been successful in reproducing many observations including the luminosity and stellar mass function of galaxies, the number counts of submillimeter galaxies galaxies, the evolution of LAEs and LBGs, the HI and HII mass functions and the AGN diversity and evolution (\citealt{Baugh2005}; \citealt{Bower2006};  \citealt{Orsi2008}; \citealt{Kim2011}; \citealt{Lacey2011}; \citealt{Lagos2011}; \citealt{Lagos2012}; \citealt{Fanidakis2011};  \citealt{Fanidakis2012}; \citealt{Gonzalez-Perez2013}). 

For the purposes of this analysis, we couple \galform with the AGN model described in \citet{Fanidakis2012}. The \citeauthor{Fanidakis2012} model follows the mass accretion rate onto the BHs and the evolution of the BH mass, $M_{\mathrm{BH}}$, and spin, $a$, allowing the calculation of a plethora of predictions related to the nature of AGN. The evolution of BHs and their host galaxies is fully coupled: BHs grow during the different stages of the evolution of the host by accreting cold (merger/disk-instability driven accretion: \emph{starburst mode}) and hot gas (diffuse halo cooling driven accretion: \emph{hot-halo mode}) and by merging with other BHs. These processes build up the mass and spin of the BH, and the resulting accretion power can regulate the gas cooling and subsequent star formation in the galaxy. The resulting mass of the BH correlates with the mass of the galaxy bulge in agreement with the observations \citep{Haering2004}. 

The gas accreted during the starburst is converted into an accretion rate, $\dot{M}$, by assuming that the accretion duration is proportional to the dynamical timescale of the host spheroid. In the hot-halo mode the accretion rate is calculated using the timestep over which gas in accreted from the halo. The bolometric luminosity of the accretion flow, $\Lbol$, is then calculated by coupling the accretion rate with the Shakura-Sunyaev thin disk \citep{Shakura1973}
\begin{equation}
\Lbol = \epsilon\dot{M}c^2,
\end{equation}
for accretion rates higher than $1$ percent of the Eddington accretion rate ($\dot{m}=\dot{M}/\dot{M}_{\rm Edd}\geqslant 0.01$) or, otherwise, the ADAF thick disk solution is adopted (\citealt{Narayan1994}; \citealt{Mahadevan1997}),
\begin{eqnarray}
L_{\mathrm{bol,ADAF}}=0.44\left(\frac{\dot{m}}{0.01}\right)\epsilon\dot{M}c^2.
\label{disc_bol_adaf}
\end{eqnarray}
When the accretion becomes substantially super-Eddington ($\Lbol\geqslant\eta\Ledd$), the bolometric luminosity is limited to (Shakura \& Sunyaev 1973)
\begin{equation}
\Lbol(\geqslant\eta\Ledd)=\eta[1+\ln(\dot{m}/\eta)]L_{\mathrm{Edd}},
\end{equation}
where $\eta$ is an ad hoc parameter, which we choose to be equal to $4$, this allows a better reproduction of the bright end of the quasar luminosity function. However, we do not restrict the accretion rate if the flow becomes super-Eddington. The luminosity output calculated via Eqns. (1), (2) and (3) is assumed to be constant during the accretion of gas With these expressions for the bolometric luminosity of the accretion flow, we henceforth define an active galaxy in our simulation to be a quasar (QSO) if its central engine exceeds $10^{46}\ergsec$ in bolometric luminosity. Active galaxies with lower bolometric luminosities will be generally described as AGN.

The merger trees of DM structures are extracted from the DM only N-body simulation MillGas (Thomas \etal in prep.). The MillGas simulation has the same mass resolution, particle number and box size as the Millennium simulation \citep{Springel2005} and differs only in the background cosmology (which is in agreement with WMAP7 results). To avoid resolution biases in the galaxy properties we calculate we consider only DM haloes with masses greater or equal to $10^{11}\Msun$. To test that DM haloes with masses lower than $10^{11}\Msun$ do not affect our predictions, we compare our findings with a simulation built on Monte-Carlo generated trees. The Monte Carlo algorithm we use to generate the DM halo merger trees has been presented in \citet{Parkinson2008}. We find no difference between the predictions of the N-body and Monte-Carlo based simulations on the galaxy properties presented in this analysis.

We note that, the AGN and galaxy formation model presented in this analysis is built upon a WMAP7 cosmology, whereas the cosmology of the original \citeauthor{Fanidakis2012} model is that of WMAP1. A change in cosmology requires retuning of the model mainly due to the change in the value of $\sigma_{8}$ from WMAP1 to WMAP7. To this end, we have retuned the model to match key BH observables in the local Universe (BH scaling relations) and the overall evolution of AGN in the $z=0-6$ Universe. These predictions, along with the properties of the background galaxy formation model, will be presented in forthcoming publications (Fanidakis \etal in prep; Lacey \etal in prep.). We refer the reader to \citet{Guo2013} for a recent discussion on how the formation, evolution and clustering of galaxies varies with cosmological parameters. 

\section{The environment of luminous quasars}

In this section we show predictions for the DM halo mass of the AGN in our model and emphasise on the host DM halo properties of the most luminous quasars ($\Lbol\geq10^{46}\ergsec$). We also calculate the expected DM halo mass of quasars in order to provide a more statistical measure for the quasar environment, one that can be directly compared with the observations.

\subsection{Distribution of AGN on the $\Lbol-\Mhalo$ plane}

\begin{figure*}
\center
\includegraphics[scale=0.86]{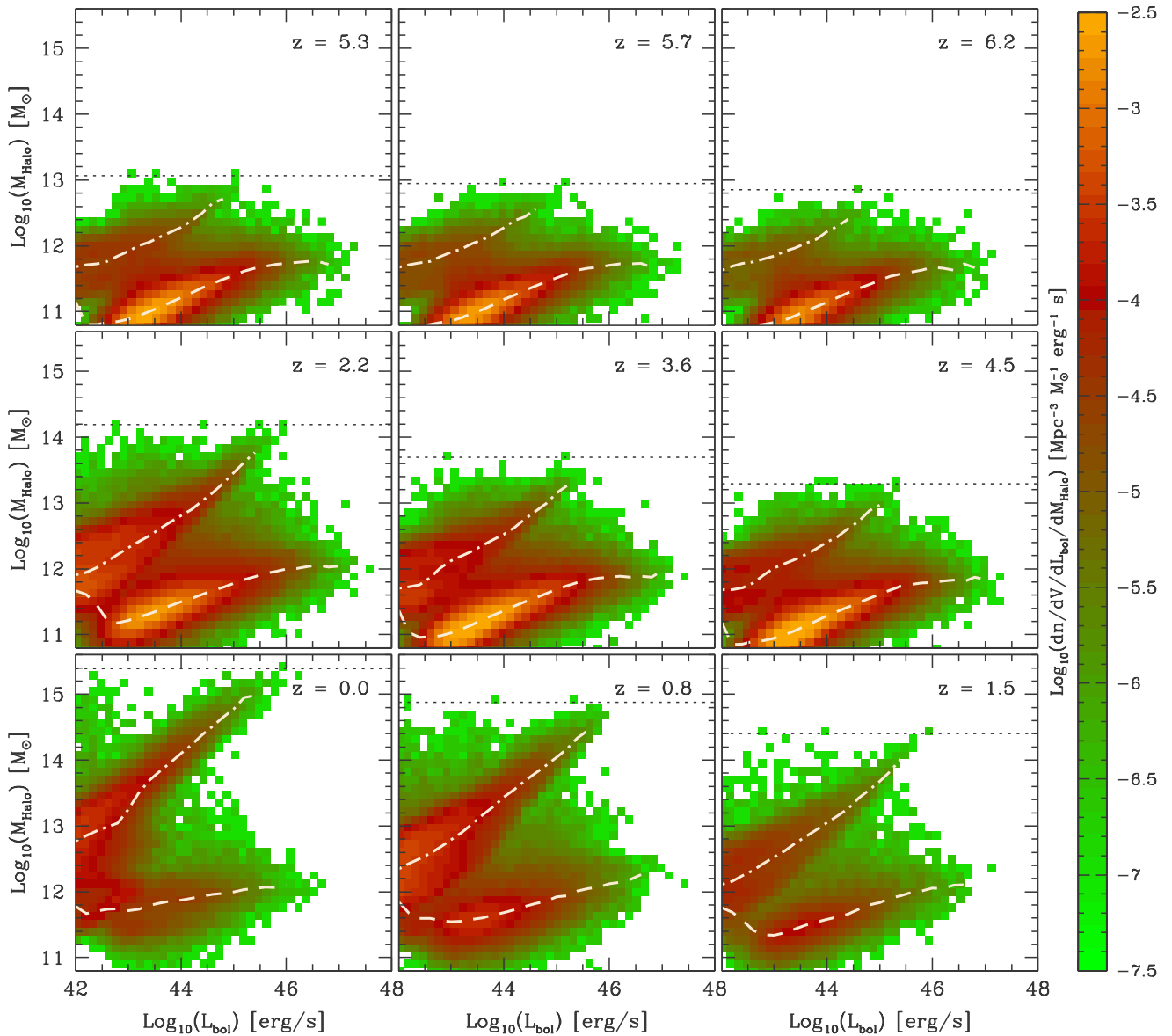}
\caption{The correlation between bolometric luminosity and DM halo mass at $z=0-6.2$ for the active galaxies in our model. Galaxies on the $\Lbol-\Mhalo$ plane are volume weighted as indicated by the colour bar. The horizontal dotted lines represent the mass of the most massive halo present in the simulation at that redshift. The white dashed and dashed-dotted lines in each panel show the median of the  $\Lbol-\Mhalo$ correlation for the starburst and hot-halo modes separately.} 
\label{mhalo_lbol}
\end{figure*}

\begin{figure*}
\center
\includegraphics[scale=0.86]{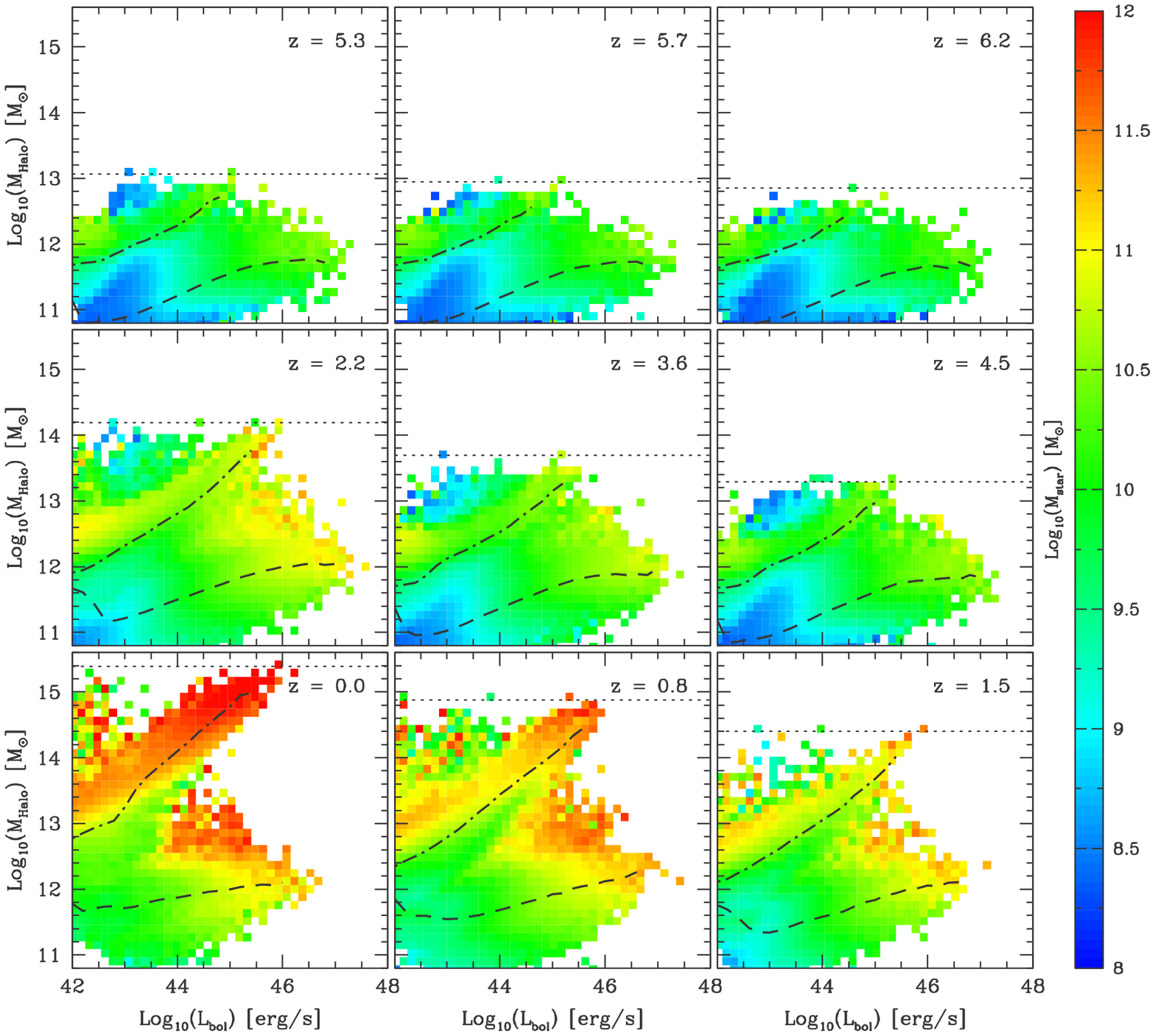}
\caption{Same as in Fig.~1, with the colour shading now representing the average host stellar mass (in bins of $M_{{\rm star}}$), as indicated by the colour bar. The horizontal dotted lines represent the mass of the most massive halo at that redshift. The dashed and dashed-dotted lines in every panel show the median of the  $\Lbol-\Mhalo$ correlation for the starburst and hot-halo modes seperately.} 
\label{mhalo_lbol_mstars}
\end{figure*}

The existence of two modes of accretion in our model  (i.e., starburst and hot-halo mode) leads to a complicated environmental dependence of AGN. This is  illustrated in Fig.~\ref{mhalo_lbol}, where we show how the bolometric luminosity correlates with DM halo mass at $z=0-6.2$. We also show the median of the $\Lbol-\Mhalo$ correlation in the starburst and hot-halo modes (dashed and dashed-dotted lines respectively) to help the reader distinguish the locus of each mode.

As illustrated by all redshift panels, AGN have a richly varied distribution on the $\Lbol-\Mhalo$ plane. Depending on the mode they accrete in, they are either found on the lower-middle part of the plane (starburst mode) or distributed diagonally upwards along the plane (hot-halo mode). In the starburst mode, the bulk of AGN is found in $\sim10^{11}-10^{12}\Msun$ haloes, although there is a very large scatter. In contrast, in the hot-halo mode we find a strong (positive) correlation between halo mass and luminosity, which extends to halo masses of $\sim10^{15}\Msun$. The shape of the two regimes remains the same with increasing redshift, but the relative density of AGN in the hot-halo mode changes significantly. At very high redshifts ($z\gtrsim5$), the hot-halo branch almost vanishes, which is mainly due to the fact that not many DM haloes are subject to AGN feedback in the early universe (as we will describe in detail in the next sections). In contrast, the starburst mode becomes the dominant mode at high redshift, mostly due to the higher abundance of cold gas in galaxies in the early universe. 

The typical stellar mass of AGN hosts varies strongly too. We show this in Fig.~\ref{mhalo_lbol_mstars}, where we now weight objects on the  $\Lbol-\Mhalo$ plane according to their stellar mass, $M_{{\rm star}}$ (indicated by the colour bar on the right). As illustrated by the individual redshift bins, in the starburst mode we find a strong correlation of luminosity with stellar mass at $z<1$, which spans approximately three orders of magnitude in stellar mass. We also find that the brightest objects, \ie the quasars, live in very massive systems with typical stellar masses of $M_{{\rm star}}\gtrsim10^{11}\Msun$. These objects are the remnants of massive disk galaxies that have recently experienced a disk instability, or occasionally a galaxy merger, and vast amounts of gas have become available for growing their central BH. Morphologically, these systems are spheroid dominated and tend to be oversized in stellar mass for their halo mass, as they represent the extreme scatter of the $\Mhalo-M_{{\rm star}}$ relation (\citealt{Moster2010}, \citealt{Moster2013}). As they are characterised by high star-formation rates (triggered during the merger or disk-instability) they \emph{are not} associated with elliptical galaxies yet. The average halo mass of these very luminous quasars remains close to $\sim10^{12}\Msun$ (with some considerable scatter), much lower than the typical halo masses of lower luminosity AGN.

For the AGN feedback to switch on in our model it is necessary for the host halo to be in hydrostatic equilibrium. That is, the cooling time of the hot gas is much longer than a multiple of the free-fall time of the halo. In addition, the BH at the centre of the halo needs to be massive enough to efficiently heat the gas in the halo via the jet \citep[see][for the details of the AGN feedback mechanism]{Bower2006}. This typically occurs at a halo mass of $10^{12}-10^{12.5}\Msun$ (the precise mass is controlled by a model parameter) and a BH mass of $10^{8.5}-10^{9}\Msun$. The most luminous quasars satisfy these conditions. Therefore, it is expected that $M_{{\rm star}}\sim 10^{11}\Msun$ galaxies that undergo a significant quasar phase soon become subject to AGN feedback. Once this happens, the central BH accretes via the hot-halo mode, which is characterised by much lower accretion rates than the starburst mode. A fraction of the accretion luminosity produced during the hot-halo mode is coupled directly to the host halo and via the AGN feedback mechanism suppresses the cooling of cold gas.

The hosts of AGN in the hot-halo branch are usually very massive in stellar mass. They live in haloes where gas cooling and star formation has been shut off by AGN feedback, and hence are red and dead. We associate these objects with the population of elliptical galaxies. Their central BHs accrete gas from the hot-halo, and due to their low density, the bolometric luminosity of the accretion flow remains low (the geometry of the flow is usually that of an ADAF). Therefore, the majority of objects in this mode have a moderate luminosity output, except from those in haloes of $\sim10^{15}\Msun$, where the central BH can shine as bright as $10^{46}\ergsec$. 

Finally, from Fig.~\ref{mhalo_lbol} we see that at $z\lesssim1.5$ the number density of objects accreting during the hot-halo mode is very high. Therefore, we expect these objects to influence strongly the typical environment of the $10^{44}-10^{46}\ergsec$ AGN. We explore this topic in \citet{Fanidakis2013} where we find that our predictions for the typical DM halo mass of moderate luminosity AGN is much higher than that of luminous quasars. This is in excellent agreement with clustering studies of moderate luminosity X-ray selected AGN (\citealt{Coil2009}; \citealt{Gilli2009}; \citealt{Cappelluti2010}; \citealt{Mountrichas2012}; \citealt{Krumpe2010}; \citealt{Starikova2011}; \citealt{Krumpe2012}; \citealt{Allevato2011}, see also \citealt{Koutoulidis2013}). 

\subsection{The effective halo mass of quasars}

We will now try to quantify the environment of quasars in a more statistical way by calculating the effective halo mass, $\Mhaloeff$, for all quasars in our simulation. $\Mhaloeff$ can be determined by an effective bias defined as \citep{Baugh1999}: 
\begin{equation}
b_{\rm eff} = \frac{\int b(\Mhalo)N_{\rm q}(\Mhalo )n(\Mhalo ){\rm d}\log \Mhalo  }{  \int N_{\rm q}(\Mhalo )n(\Mhalo ){\rm d}\log \Mhalo},
\end{equation}
where $b$ is the bias of DM haloes with mass $\Mhalo$, $N_{\rm q}(\Mhalo )$ is the mean number of quasars in a halo of mass $\Mhalo$ and $n(\Mhalo )$ is the number density of DM haloes with mass $\Mhalo$. We calculate $\Mhaloeff$ from $b_{\rm eff}$ using the ellipsoidal collapse model of \citep{Sheth1999}. Therefore the model predictions can be compared directly to the observational estimation of quasar halo masses from surveys such as SDSS and 2dF.  

\begin{figure}
\center
\includegraphics[scale=0.43]{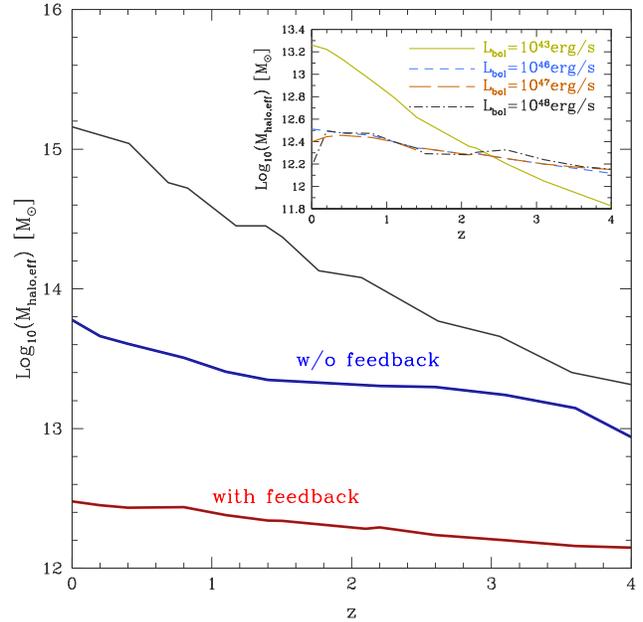}
\caption{The effective DM halo mass of quasars ($\Lbol\geq10^{46}\ergsec$) as a function of redshift  (solid red line). Predictions are also shown for a simulation where AGN feedback is not taken into account (blue solid line). The grey solid line represents the mass of the most massive halo in the simulation at a given redshift. Inset panel: Effective DM halo mass as a function of redshift for four different luminosity AGN populations,  as indicated by the key (quasars represent the $\Lbol= 10^{46}, 10^{47}$ and $10^{48}\ergsec$ populations).} 
\label{effective_mhalo}
\end{figure}

The $\Mhaloeff$ of quasars as a function of redshift is shown in Fig.~\ref{effective_mhalo} (solid red line). As illustrated by the plot, $\Mhaloeff$ remains close to $\sim10^{12.4}\Msun$ for quasars in the low-redshift universe, and drops by $\sim0.2\,\rm{dex}$ as $z$ increases. This prediction is consistent with what clustering analyses of quasars in galaxy surveys indicate (\citealt{Croom2005}; \citealt{Ross2009}), namely that quasars in the low-$z$ Universe tend to live in average DM environments. We note that, to produce such high luminosities as $10^{48}\ergsec$, a vast amount of gas is required to be accreted, which is why we find our brightest sources in the most gas rich (and massive) galaxies. Yet, even these extreme objects tend to be found in haloes of similar mass ($\sim10^{12}\Msun$) to the hosts of average quasars. We show this in the inset panel of  Fig.~\ref{effective_mhalo}, where we plot $\Mhaloeff$ as a function of redshift for different luminosity AGN populations ($\Lbol= 10^{43},10^{46}, 10^{47}$ and $10^{48}\ergsec$). Quasars ($\Lbol>10^{46}\ergsec$) are found to have the same environment, with only insignificant variations at low redshift. In contrast, AGN belonging to much less luminous classes (e.g., $\Lbol=10^{43}\ergsec$) tend to live in haloes more massive by an order of magnitude at low redshift.

Finally, in \citet{Fanidakis2013}, we suggest that the luminosity output of an AGN is determined by the accretion channel and ultimately by the DM halo mass of the AGN host. For example, haloes that are subject to AGN feedback can only host AGN whose accretion flow is relatively under dense, and therefore produce moderate luminosities ($10^{44}-10^{46}\ergsec$). Accretion in this case is fed directly by the hot halo around the galaxy. In this picture, quasars can only exist in average environments where AGN feedback is not present and thus, gas in the host halo can cool efficiently. The importance of AGN feedback in defining the halo mass of bright QSO, is illustrated by the blue line in the main panel of Fig.~\ref{effective_mhalo}, which shows $\Mhaloeff$ for quasars in a simulation where AGN feedback is turned off. In this case, the typical halo mass of quasars\footnote{In this simulation we consider all AGN with $\Lbol\gtrsim10^{46}\ergsec$ as quasars. We note, that in this case, the model has not been tuned to fit the observational data. Requirement of retuning the accretion model might result in lower AGN luminosities, however, without affecting the clustering of these sources much.} increases dramatically, making their environment that of the very massive haloes. In fact, the largest halo at a given redshift (indicated by the solid grey line in the plot) is now found to host a very bright quasar with typical luminosity of $\sim10^{47}-10^{48}\ergsec$. Hence, in a universe without AGN feedback, galaxy groups and clusters should be the typical environments where enormous quasar activity takes place. However, in our observable Universe bright quasars are never observed in the cores of clusters in the low-redshift Universe.

\section{The environment of first quasars}

As mentioned earlier, $z\sim6$ quasars are of particular interest, since they  are assumed to reside in overdensities and pinpoint the location of protoclusters. Here we test this idea by considering the $z\sim6$ quasars in our model and exploring the properties of their DM halo environment and their descendants at $z=0$. We also provide a calculation for the number of LBGs expected to be found around $z\sim5$ quasars in order to reconcile the observations that search for galaxy overdensities around high-$z$ quasars. 

\begin{figure}
\center
\includegraphics[scale=0.43]{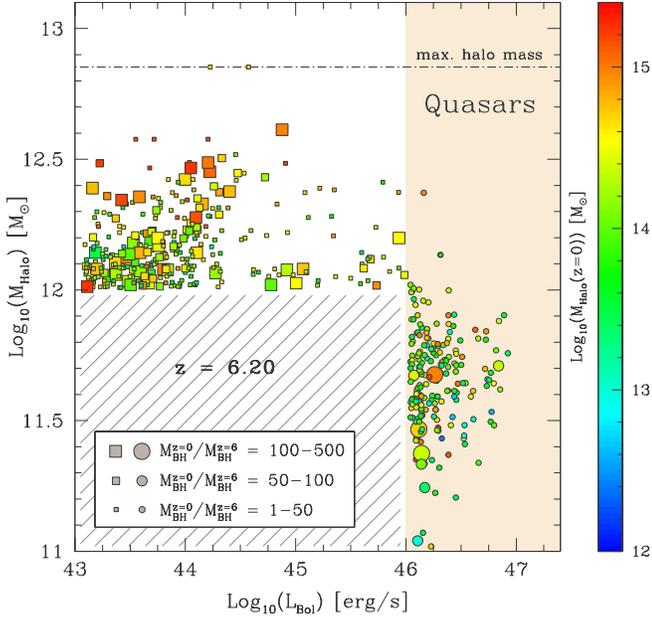}
\caption{Bolometric luminosities and DM halo masses for $z=6.2$ quasars (filled circles). We also plot all haloes with masses greater than $10^{12}\Msun$ that host an AGN  (filled squares). Symbols are colour coded according to the halo mass of the $z=0$ descendant, as indicated by the colour bar on the right. Symbol sizes indicate the ratio of the quasar descendant BH mass at $z=0$ over its BH mass at $z=6.2$. The dashed-dotted horizontal line indicates the mass of the most massive halo at $z=6.2$. The hatched area represents the part of the plane that is not sampled for clarity reasons.} 
\label{first_quasars}
\end{figure}

\subsection{The $z=6.2$ quasars and their descendants to $z=0$}

In this section, we explore the environment of high-$z$ quasars and present predictions for their descendants at $z=0$. We do so by showing in Fig.~\ref{first_quasars} how quasars (filled circles) populate the $\Lbol-\Mhalo$ plane at $z=6.2$. In addition, we plot all DM haloes (filled squares) with $\Mhalo>10^{12}\Msun$ that host an AGN. We sample these objects from a subvolume of $200\,{\rm Mpc}/h$ of the simulation to avoid overfilling the plot (yet the volume is large enough to exclude cosmic variance effects). Objects on the $\Lbol-\Mhalo$ plane are colour coded according to their descendant halo mass at $z=0$, as indicated by the colour bar on the right. In addition, the size of the symbols indicates the ratio of the central BH mass at $z=0$ to that at $z=6.2$

According to Fig.~\ref{first_quasars}, our model suggests that the halo hosts of luminous quasars at $z=6.2$ span an order of magnitude in mass, which always remains between $\sim10^{11}-10^{12}\Msun$ (with only one or two exceptions higher than $10^{12}\Msun$). When considering the extremes of the quasar population, we find that the brightest quasar at $z=6.2$ has a bolometric luminosity of $10^{46.9}\ergsec$ and a host halo mass of $\sim10^{11.8}\Msun$. Its descendant at $z=0$ is a central elliptical galaxy with stellar mass of $\sim10^{11.4}\Msun$ in a DM halo of mass $10^{13.4}\Msun$. Interestingly, the BH harboured by this quasar has grown by a factor of six in mass by $z=0$. 

The galaxy descendants of the rest of the $z=6.2$ quasars show a wide range of morphologies. Even though the majority of them ($60$ percent) evolve in to pure spheroids, we find a non-negligible fraction ($15$ percent) of disk galaxies ($B/T\lesssim0.6$). These are usually satellites galaxies in a variety of halos (frequently also in $10^{15}\Msun$ haloes) with relatively low stellar masses ($10^{12}-10^{11}\Msun$). Another interesting aspect of these galaxies is that their central BHs have not grown much since $z\sim6$. The characteristics of pure spheroidal descendants (which at $z=0$ are elliptical galaxies) are also quite diverse, although they do tend to be more massive ($\gtrsim10^{11}\Msun$) and centrals. In these galaxies we find that BHs usually grow more efficiently and, as seen in Fig.~\ref{first_quasars}, in some of the descendants the central BH has grown by more than 2 orders of magnitude in mass since $z\sim6$. Interestingly, we find no apparent correlation between $z=6.2$ quasar luminosity and descendant halo mass, stellar mass or morphology. 

Similarly to the low-$z$ universe, the most massive haloes tend to avoid hosting a quasar. This result is in contrast with what is usually assumed, as outlined previously. A great fraction of haloes at $z=6$ with masses higher than $10^{12}\Msun$ is found to host an AGN of moderate luminosity. The most massive halo at $z=6.2$ ($\Mhalo=10^{12.85}$\Msun) does host an AGN, its luminosity though is relatively low and equal to $\sim10^{44.6}\ergsec$, which makes it too faint to be detected with current instruments. The halo descendant of the most massive halo at $z = 6$ has a mass of $10^{14.7}\Msun$ and hosts a massive elliptical with $M_{{rm star}}=10^{11.5}\Msun$. Interestingly, this is not the most massive DM halo in our cosmological volume at  $z=0$. This is in agreement with recent findings by \citet{Angulo2012}. These authors showed that in a hierarchical universe, the future growth of a $z\sim6$ DM halo is more determined by the environment on scales of $\sim10\,{\rm Mpc}$ than by the actual halo mass, as the former is expected to be the main factor shaping the future halo mass assembly.

\begin{figure*}
\center
\includegraphics[scale=0.43]{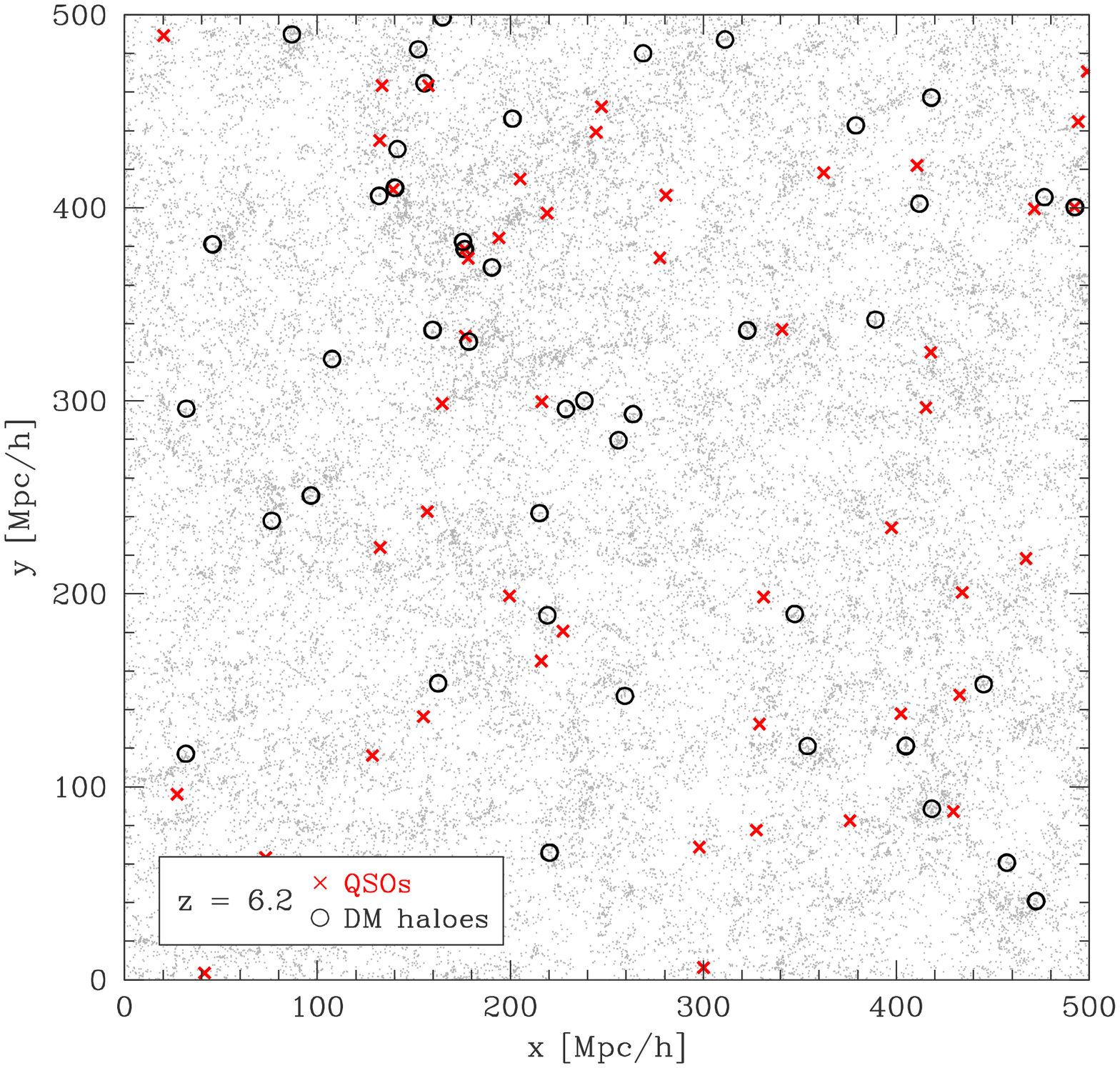}
\includegraphics[scale=0.43]{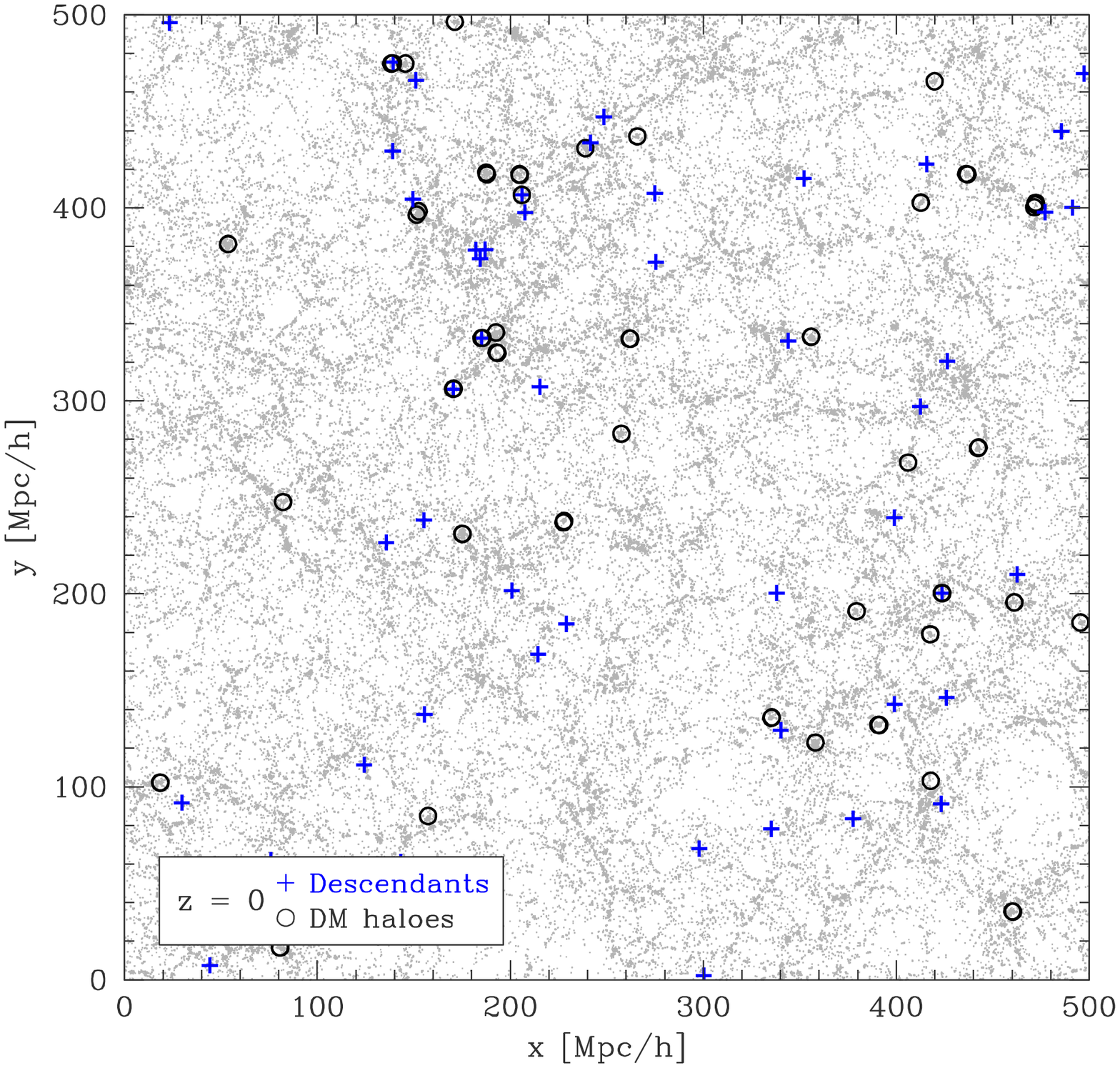}
\caption{Spatial maps of QSOs and the most massive DM haloes. Left: The spatial distribution of QSOs (red crosses) and the most massive DM haloes (black circles) at $z=6.2$ in our simulation. Right: The spatial distribution of the descendant DM haloes of $z=6.2$ QSOs (blue crosses) and the most massive haloes at $z=0$. The thickness of each slice is $100\,h^{-1}{\rm Mpc}$ along the ${\rm z}$ direction. In both panels, the number of DM haloes depicted equals the number of QSOs. The underlying DM distribution is shown in grey.} 
\label{first_quasars_maps}
\end{figure*}

The lack of quasars in the most massive DM haloes in the early universe is due to the universality of the AGN feedback mechanism. Even at such high redshifts, haloes with masses greater than $\sim10^{12}\Msun$ become subject to feedback. Their space density is very low ($<10^{-8}\,\rm{Mpc^{-3}}$), nevertheless, as we shall show in a forthcoming study, BH growth and quasar activity in these environments is considerably affected by AGN feedback. Similar conclusions have been reached by \citet[][see also \citealt{Degraf2012}]{DiMatteo2012}. These authors have employed high resolution SPH simulations of the growth of BHs in the early universe (with a box size of $0.75\,h^{-1}{\rm Gpc}$) to show that the most massive haloes ($\Mhalo\gtrsim10^{12}\Msun$) by $z=6$ have already shut off quasar activity at their centres due to feedback.


The $z=0$ halo descendants of luminous quasars are characterised by a wide range of masses, typically higher than $10^{13}\Msun$. In particular, we find that $42$ percent of the $z=6.2$ quasars have a halo descendant with mass in the range $10^{13}-10^{14}\Msun$, $48$ percent have a mass of $10^{14}-10^{15}\Msun$ (with the majority of them close to $\sim10^{14}$), and $14$ percent have a halo descendant more massive than $10^{15}\Msun$. Interestingly, when we consider quasar descendants that are central haloes we find that the fraction in the $10^{14}-10^{15}\Msun$ mass regime decreases to only $15$ percent. Also, we do not find central haloes more massive than $10^{15}\Msun$. This means that a great fraction of these descendants evolve only to become satellites of the most massive haloes at $z=0$. Hence, the majority of $z\sim6$ quasars although they have possibly been close to the actual progenitors of the $z=0$ most massive haloes, they do not really coincide with them at that time. We note that the quasar hosts that evolve to $z=0$ central haloes with mass of $10^{14}-10^{15}\Msun$ represent a fraction of only $0.6$ percent of the total halo population at $z=0$ with such mass. Thus, the picture that emerges when comparing $z\sim6$ quasar host haloes and their descendants at $z=0$ is the following. The present-day $\geq10^{15}\Msun$ haloes did not host a quasar at $z=6.2$ (although there is a non negligible probability of 0.14 that a quasar was in their near vicinity), while there is a probability of $0.15$\footnote{The probability of 0.15 (which arises from the fact that 15 percent of quasar halo descendants evolve to $10^{14}-10^{15}\Msun$ haloes at $z=0$) is really an upper limit since these quasar haloes could have been satellites at high redshift that merged with the central object before $z=0$.} that the halo descendant of a $z=6$ quasars will coincide with a $10^{14}-10^{15}\Msun$ halo at $z=0$. Thus, we can conclude that $z\sim6$ quasars could possibly trace the progenitors of rich clusters ($\Mhalo\sim10^{14}\Msun$) but not superclusters ($\Mhalo\sim10^{15}\Msun$). 

The properties of quasars on the $\Lbol-\Mhalo$ plane in Fig.~\ref{first_quasars} are confirmed also when we consider how quasars are distributed in the cosmic web. We show this in Fig.~\ref{first_quasars_maps}, where we plot the spatial distribution of quasars relative to that of the most massive DM haloes at $z=6.2$ (left panel) and the distribution of their descendants at $z=0$ (right panel). At $z=6.2$ we find, that in the majority of the cases quasars avoid the environments of the extreme DM peaks. In cases where the position of quasars coincides with that of the largest DM haloes, we find that the host halo has a mass of $\sim10^{12}\Msun$, in accordance with Fig.~\ref{first_quasars}. At $z=0$ we find that the quasar descendants are found in less extreme environments of the DM distribution, yet occasionally also in the most massive haloes. This picture illustrates again that some of the most massive DM haloes at $z=0$ could have hosted a bright quasar at $z=6.2$. However, detecting a bright quasar at $z=6.2$ does not guarantee that its host halo is a progenitor of a massive halo at $z=0$.

\subsection{Overdensities around $z\sim5$ quasars}

As we saw previously, the halo mass of low-$z$ quasars is well below the mass of the most massive structures of the DM distribution. For the majority of quasars their environment is representative of that of the average mass DM haloes. However, at $z\sim6$ the mass of the most massive haloes is only an order of magnitude higher than the typical halo mass of quasars. Therefore, we should expect an enhancement of structures near these quasars (especially for those hosted by $10^{12}\Msun$ haloes), yet not as strong as the overdensities in the most massive haloes. 

\citet{Husband2013} recently presented an observational study of the clustering of LBGs in three quasar fields at $z\sim5$. These authors employed spectroscopically identified LBGs in the ESO Remote Survey (ERGS; Douglas et al 2009, 2010) and showed that two of the three fields show an overdensity of galaxies within a narrow redshift range of $\Delta z=0.05$. When comparing to the clustered structures identified in ERGS, the authors conclude that QSO environments are overdense, yet not more extreme than rich structures in the field. Here, we test this observation by calculating the number of LBGs expected to be found near a quasar in our simulation. 

\begin{figure}
\center
\includegraphics[scale=0.42]{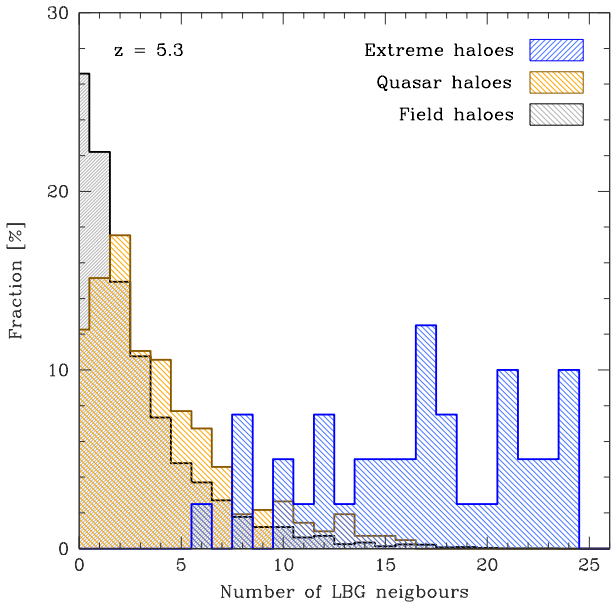}
\caption{The fraction of DM haloes at $z=5.3$ hosting $n$ LBGs within a radius of $2$\,Mpc. The different histograms represent the haloes that host quasars (orange shading), the most massive DM haloes ($\Mhalo\geq5\times10^{12}\Msun$, blue shading) and the ``field" DM haloes ($\Mhalo\geq10^{11}\Msun$, grey shading). The lower luminosity cut for the LBG sample is $M_{\rm AB}(1500\rm{\AA})=-16$.} 
\label{lbg_numcounts}
\end{figure}

We model LBGs as in \citet{Lacey2011}, by taking into account the attenuation of starlight by the dust content of the galaxy. The predictions of the model for the abundance of LBGs matches the observed LBG LF over a wide range of redshifts ($3<z<10$) and the clustering of LBGs at $3\lesssim z\lesssim6$ (see Lacey et al., in prep). Our sample of LBGs considers all galaxies with far-UV ($1500{\rm \AA}$) luminosities in the range determined by the LBG luminosity function at $z\sim5$, i.e. $M_{\rm AB}(1500\rm{\AA})\leq-16$. With this luminosity cut, the model predicts an LBG space density of $1.3\times10^{-3}{\rm Mpc}^{-3}$. We determine the number of expected LBG neighbours by counting the total number of LBGs within a sphere of radius $2$\,Mpc centred around the quasar and normalise it in units of the mean. Our predictions at $z= 5.3$ are shown in Fig.~\ref{lbg_numcounts}. We also show the expected number of LBGs in the field and in the most massive (extreme) DM structures by selecting all DM haloes with $\Mhalo>10^{11}\Msun$ and $\Mhalo>5\times10^{12}\Msun$ respectively. We do not consider haloes less massive than $10^{11}\Msun$ in the calculation of the field abundance due to the low resolution of the simulation below that mass.   

According to Fig.~\ref{lbg_numcounts}, our model suggests that the number of LBGs expected to be found within $2$\,Mpc from a quasar is similar to that of the field. However, $n\geq2$ neighbours are expected to be found in a higher fraction of quasars than in field DM haloes. This implies that an overdensity is more likely to be found around a quasar than in the field, in good agreement with the analysis of Husband et al. On the other hand, LBGs cluster strongly near the most-massive haloes, where we find that the typical number of neighbours is $\sim10-25$ galaxies. Thus, we expect a considerable number of LBGs at the extremes of the DM distribution, much higher than the one found around quasars. Thus although an enhancement of galaxies is expected to be found around quasars (with a frequency higher than when searching blank fields), the number of galaxies detected is considerably lower than that expected in the most massive structures.

We note that, our predictions are very sensitive to the luminosity cut of the LBG sample. Surveys that do not reach magnitudes as faint as the ones we consider here will of course detect a lower number of objects. When considering brighter samples we find an overall decrease in the number of LBGs around quasars, field and extreme haloes, yet the overall picture does not change much. Hence we conclude, although it is very likely to find an overdensity of galaxies around a quasar compared to the field, the actual extremes of the DM distribution show a much stronger clustering signal compared to the environments traced by quasars.

\section{Summary-Conclusions}

We have presented an analysis of the DM halo environments of quasars (active galaxies with bolometric nuclei luminosities greater of $10^{46}\ergsec$) using the semi-analytic model \texttt{GALFORM}. We have found that quasars live in average environments with a typical halo mass of $10^{12}\Msun$. This halo mass remains approximately constant up to $z\sim4$ and is insignificantly dependent on luminosity. The triggering of quasar activity in higher mass haloes is usually inhibited by the AGN feedback mechanism. When switching off feedback in our calculation, we find a typical halo mass higher than $10^{13}\Msun$. This is in contrast with the observational estimates from SDSS and 2dF. Therefore, AGN feedback is necessary not only for reproducing the right shape of the galaxy luminosity function, but also the halo environment of quasars. 

At higher redshifts quasars reside in massive haloes that have not yet become subject to feedback. At $z\sim6$ these haloes have masses of $10^{11}-10^{12}\Msun$, almost an order of magnitude lower than the mass of the most massive bound structures at that redshift. The descendants of these haloes span a wide range of halo masses, stellar content and morphologies. We find that an important fraction ($15$ percent) of these quasars have evolved into disk satellites in massive haloes ($\Mhalo=10^{12}-10^{13}\Msun$), with their central BHs having grown only by a factor of a few since $z=6$. Quasars also evolve into spheroidal galaxies (which are associated with elliptical galaxies), typically found in rich clusters ($\Mhalo\sim10^{14}\Msun$). When we consider $z\sim6$ as cosmological probes of protoclusters, we find that a small fraction ($15$ percent) of them could be linked to the progenitor of $z=0$ rich clusters. As far as the $z=0$ superclusters are concerned ($\Mhalo\sim10^{15}\Msun$), the model predicts that these haloes did not host a quasar at $z\sim6$. Therefore, $z\sim6$ quasars do not trace the progenitors of the present-day superclusters. 

Regarding the abundance of galaxies detected around high-$z$ quasars, we find that when searching the fields of $z\sim5$ quasars, it is very likely to find an overdensity of galaxies. In fact, the probability of finding one is higher than when searching blank fields. However, the overdensities detected around quasars are not as extreme as those expected in the most massive haloes at that redshift. Therefore, observations that find overdensities around quasars do not really probe the peaks of the DM distribution. The model also predicts a significant number of quasars that do not reside in an overdensity. The same picture seems to hold also at $z\sim6$.

To conclude, we have shown that the halo mass of quasars at all redshifts is average and does not coincide with that of the most massive structures of the DM distribution. This study is a first step towards understanding the environment of quasars in the low and high-$z$ universe. In a forthcoming paper we will explore in detail the complete evolution through cosmic time of the first quasars in order to understand better the nature of the most luminous objects in the Universe. 

\section*{Acknowledgements}

The authors would like to thank Raul Angulo and Aaron Dutton for valuable comments. AVM acknowledges financial support to the DAGAL network from the People Programme (Marie Curie Actions) of the European Union’s Seventh Framework Programme FP7/2007- 2013/ under REA grant agreement number PITN-GA-2011-289313. CMB acknowledges Max-Planck-Institut f\"ur Astronomie for its hospitality and financial support through the Sonderforschungsbereich SFB 881 ``The Milky Way System'' (subproject A1) of the German Research Foundation (DFG).


\bibliographystyle{mn2e}
\bibliography{fanidakis_2013b}

\end{document}